\documentclass[twocolumn]{aa}
\usepackage{graphicx,epsf}
\usepackage[figuresright]{rotating}
\begin{document}
\renewcommand{\topfraction}{0.85}
\renewcommand{\bottomfraction}{0.7}
\renewcommand{\textfraction}{0.15}
\renewcommand{\floatpagefraction}{0.90}
   \title{Evidence for VHE $\gamma$-ray Emission from the Distant BL Lac PG\,1553+113}

\author{F. Aharonian\inst{1}
 \and A.G.~Akhperjanian \inst{2}
 \and A.R.~Bazer-Bachi \inst{3}
 \and M.~Beilicke \inst{4}
 \and W.~Benbow \inst{1}
 \and D.~Berge \inst{1}
 \and K.~Bernl\"ohr \inst{1,5}
 \and C.~Boisson \inst{6}
 \and O.~Bolz \inst{1}
 \and V.~Borrel \inst{3}
% ...
}

  \offprints{Wystan.Benbow@mpi-hd.mpg.de or santiago.pita@cdf.in2p3.fr}
 
\institute{
Max-Planck-Institut f\"ur Kernphysik, Heidelberg, Germany
\and
 Yerevan Physics Institute, Armenia
\and
Centre d'Etude Spatiale des Rayonnements, CNRS/UPS, Toulouse, France
\and
Universit\"at Hamburg, Institut f\"ur Experimentalphysik, Germany
\and
Institut f\"ur Physik, Humboldt-Universit\"at zu Berlin, Germany
\and
LUTH, UMR 8102 du CNRS, Observatoire de Paris, Section de Meudon, France
% ...
}
 
   \date{Received 14 December 2005 / Accepted 19 January 2006}

   \abstract{
The high-frequency peaked BL Lac PG\,1553+113 was observed in 2005 
with the H.E.S.S. stereoscopic array of imaging atmospheric-Cherenkov telescopes in Namibia. 
Using the H.E.S.S. standard analysis, an excess was measured at the 4.0$\sigma$ level 
in these observations (7.6 hours live time).  Three alternative, lower-threshold analyses yield
$>$5$\sigma$ excesses. The observed integral flux above 200 GeV is 
(4.8$\pm$$1.3_{stat}$$\pm$$1.0_{syst}$)$\times$10$^{-12}$ cm$^{-2}$\,s$^{-1}$,
and shows no evidence for variability.  
The measured energy spectrum is characterized by a very soft power law 
(photon index of $\Gamma$=4.0$\pm$0.6).  Although the redshift of PG\,1553+113
is unknown, there are strong indications that it is greater than $z$=0.25 
and possibly larger than $z$=0.78.  The observed spectrum is interpreted in
the context of VHE $\gamma$-ray absorption by the Extragalactic Background Light, and is 
used to place an upper limit on the redshift of $z$$<$0.74.

   }

   \maketitle
%
%________________________________________________________________

\section{Introduction}

PG\,1553+113 was discovered in the Palomar-Green survey of UV-excess stellar objects 
\cite{discovery_paper}.  It is classified as a BL Lac object
based on its featureless spectrum (\cite{redshift_initial, redshift_question}) 
and its significant ($m_p$ = 13.2-15.0) optical variability \cite{optical_variability}.  
PG\,1553+113 is well studied from the radio to the X-ray regime, and has been the subject of
several multi-wavelength observation campaigns.  In X-rays it 
has been detected by multiple observatories, with energy spectra
measured by both {\it BeppoSAX} \cite{BeppoSAX_det} and {\it XMM Newton} \cite{XMM_det}.
Based on its broad-band spectral energy distribution (SED), PG\,1553+113
is now classified as a high-frequency peaked BL Lac \cite{classification}, similar 
to essentially all of the AGN detected at VHE (very high energy; $>$100 GeV) energies.

The redshift of PG\,1553+113 was initially determined to 
be $z$=$0.36$ (\cite{redshift_initial}).  However, it was 
later claimed that this value is flawed as it was based on a spurious emission line, caused by a
bright spot in the {\it International Ultraviolet Explorer} image, 
that was misidentified as Lyman-$\alpha$ (\cite{redshift_question}).
To date no emission or absorption lines have been measured despite more than ten
observation campaigns with optical instruments, including two with 
the 8-meter VLT telescopes \cite{no_lines}.  Estimates of
the redshift can also be made using photometric observations. No host galaxy was
resolved in {\it Hubble Space Telescope} (HST) images of PG\,1553+113 
taken during the HST survey of 110 BL Lac objects \cite{Hubble_image}. 
Approximately 80\% of these BL Lacs (88) have known redshifts, of which
all 39 with $z$$<$0.25 and 21 of the 28 with 0.25$<$$z$$<$0.6 
have their hosts resolved \cite{z_extreme}.  The HST results were recently used
to set a lower limit of $z$$>$0.78 \cite{z_extreme} for PG\,1553+113.
This redshift limit is based on the assumption of the nucleus
being present in a typical host galaxy, and general correlation between 
photometrically determined redshifts and those determined 
spectroscopically. However, a departure from this correlation, 
or the possibility of an atypical host galaxy is not unexpected.  The possibility of such a 
large redshift is of critical importance to VHE  observations 
due to the absorption  of VHE photons (\cite{EBL_effect, EBL_effect2}) 
by pair-production on the Extragalactic Background Light (EBL). 
This absorption, which is energy dependent and increases strongly with
redshift, distorts the VHE energy spectra observed from distant objects,
and may even render the VHE detection of very distant objects impossible.

Although the redshift of PG\,1553+113 is essentially unknown, it is viewed as a 
promising candidate for detection as a VHE emitter \cite{luigi_AGN} based on its SED.  
However, it has not been previously detected in the VHE regime. 
Only flux level upper limits have been reported by the 
Whipple collaboration \cite{whipple_UL}
and the Milagro group \cite{Milagro_AGN}.
Here, evidence for $>$200 GeV $\gamma$-ray emission from PG\,1553+113 
is presented, based on observations made with the 
H.E.S.S. (High Energy Stereoscopic System) experiment. 
The data are used to set an upper limit on the source redshift.

\section{H.E.S.S. Detector \& Observations}
The H.E.S.S. experiment is a stereoscopic system of four imaging 
atmospheric-Cherenkov telescopes located in Namibia.
H.E.S.S. is designed to search for astrophysical $\gamma$-ray emission
above $\sim$100 GeV, with a sensitivity (5$\sigma$ in 25 hours for a 1\% Crab Nebula flux
source at 20$^{\circ}$ zenith angle) that allows for the detection of
VHE emission from objects such as PG\,1553+113 at previously
undetectable flux levels.  More details on H.E.S.S. can be
found in \cite{HESS1}.

The H.E.S.S. observations of PG\,1553+113 were made in May and August 2005.  A
total of 22 runs ($\sim$28 minutes each) were taken, yielding a total exposure of 9.6 hours.
Seventeen of these runs pass selection criteria which remove data for which the weather 
conditions were poor or the hardware was not functioning properly.  The total exposure
of the quality selected data set is 7.6 hours live time, and the mean zenith angle of
these observations is 40$^{\circ}$.  All these data were taken
using {\it Wobble} mode, i.e.  the source direction is positioned $\pm$0.5$^{\circ}$ relative 
to the center of the field of view (f.o.v.) of the camera during observations, which
allows for both on-source observations and simultaneous estimation
of the background induced by charged cosmic rays.

\section{Analysis Technique}

The data passing the run selection criteria are calibrated as detailed in \cite{calib_paper},
and the event reconstruction and background rejection (i.e. event selection criteria)
are performed as described in Aharonian et al. (2005a),
with some minor improvements discussed in Aharonian et al. (2005b).
On-source data are taken from a circular region of radius $\theta_{cut}$ centered on the source.
The background is estimated using events from a number of equal-size
off-source regions offset by the same distance, relative
to the center of the f.o.v., in the sky as the on-source 
region (for more details see \cite{AGN_UL_paper}).
The significance of any excess is calculated following the method of Equation (17) in \cite{lima}.

Table~\ref{thecuts} shows the cuts on mean reduced scaled width (MRSW)
and length (MRSL) parameters \cite{pks2155_paper}, on $\theta^{2}$ (the 
square of the angular difference between the reconstructed shower direction and
the source position), on individual image size (in photoelectrons; PE), 
and on the distance of the image center of
gravity from the center of the f.o.v., used in this analysis.  
The standard H.E.S.S. analysis cuts shown in Table~\ref{thecuts} are 
optimized {\it a priori} using Monte Carlo $\gamma$-ray simulations and unrelated
off-source data to yield the maximum expected significance per hour of 
observation for a weak (10\% Crab flux source with photon index $\Gamma$=2.6) source.  
Generally, the significance expected (and observed) is not strongly dependent on the
exact values of the cuts.  However, the size cut increases the energy threshold of the
analysis, therefore lower values are more appropriate for a steep spectrum source.

   \begin{table}
      \caption{The selection cuts applied to the data for the H.E.S.S. standard analysis
and those applied for the spectrum determination in this article.  
Cuts keeping only events with MRSL and MRSW greater than $-2.0$$\sigma$ are used in both cases.
Only images passing the distance and size cuts
are used in the analysis, and a minimum of two such images are required.}
         \label{thecuts}
         \centering
         \begin{tabular}{c   c   c   c   c   c   c   c}
            \hline\hline
            \noalign{\smallskip}
            Cut & MRSL & MRSW & $\theta_{cut}^2$ & Size & Dist. \\
            Type & max & max & max & min & max\\
            & [$\sigma$] & [$\sigma$] & [deg$^{2}$] & [PE] & [mrad]\\
            \noalign{\smallskip}
            \hline
            \noalign{\smallskip}
            Standard & 2.0 & 0.9 & 0.0125 & 80 & 35\\
            Spectrum & 1.3 & 0.9 & 0.02 & 40 & 35\\
            \noalign{\smallskip}
            \hline
         \end{tabular}
   \end{table}
   
\section{Results}

   \begin{table}
      \caption{Results of the H.E.S.S. observations of  PG\,1553+113 using both the standard analysis 
	cuts and the lower-threshold spectral analysis cuts.  Shown are
        the post-cuts energy threshold at the mean zenith angle of observations (40$^{\circ}$),
	the number of on-source and off-source events passing the cuts,
        the normalization for the off-source events, the observed excess ($\Delta$) from PG\,1553+113
        and the significance (S) of the excess.  The significance resulting from the spectral analysis
	cuts is subject to a small trials factor.  Therefore only the significance resulting from the
	H.E.S.S. standard analysis should be considered.}
         \label{results}
        \centering
         \begin{tabular}{c c c c c c c}
            \hline\hline
            \noalign{\smallskip}
	    Cut & E$_{\mathrm{th}}$ & & & & & S\\
            Type & [GeV] & On & Off & Norm & $\Delta$ & [$\sigma$]\\
            \noalign{\smallskip}
            \hline
            \noalign{\smallskip}
            Standard & 310 & 457 & 4126 & 0.0907 & 83 & 4.0\\
            Spectrum & 230 & 1807 & 12753 & 0.1236 & 230 & 5.3\\
           \noalign{\smallskip}
	    \hline
       \end{tabular}
   \end{table}

\begin{figure}
   \centering
      \includegraphics[width=8.7cm]{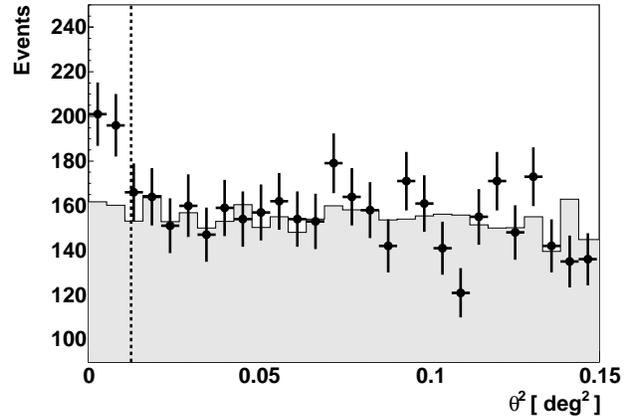} \\ [-0.3cm]
      \caption{The distribution of $\theta^2$ for on-source events (points) and
	normalized off-source events (shaded) from observations of PG\,1553+113. 
	The dashed line represents the cut on $\theta^2$ applied to the data.}
         \label{thtsq_plot}
   \end{figure}

Table~\ref{results} shows the results from the analysis of the H.E.S.S. observations
of PG\,1553+113.  Using the standard cuts, 
an excess of 83 events ($4.0$$\sigma)$ is observed from the direction of 
PG\,1553+113.  Figure~\ref{thtsq_plot} shows the on-source 
and normalized off-source distributions of $\theta^{2}$ for the H.E.S.S. observations. 
The background is approximately flat in $\theta^{2}$ as expected, and there is an excess at 
small values of $\theta^{2}$ corresponding to the observed signal. The shape of
this excess is consistent with that expected from a point source given the H.E.S.S. point-spread function.

Assuming a power-law spectrum ($dN/dE$\,$\sim$\,$E^{-\Gamma}$)
with a photon index of $\Gamma$=4.0 (justified later), 
the observed integral flux above 200 GeV is 
I($>$200 GeV)=(4.8$\pm$$1.3_{stat}$$\pm$$1.0_{syst}$)$\times$10$^{-12}$ cm$^{-2}$\,s$^{-1}$.  
This corresponds to $\sim$2\% of I($>$200 GeV) determined by H.E.S.S. for the Crab Nebula.
The flux is well below all previously published upper limits for PG\,1553+113.  
No evidence for variability is found as a fit to the nightly integral flux versus time 
is consistent with being constant ($\chi^2$=3.7, 7 degrees of freedom).
Additionally, quick-look results provided by the RXTE/ASM team show no indication
of flaring behavior at X-ray energies (2 to 10 keV), suggesting
that PG \,1553+113 was not in a high-emission state.

   \begin{figure}
   \centering
      \includegraphics[width=8.7cm]{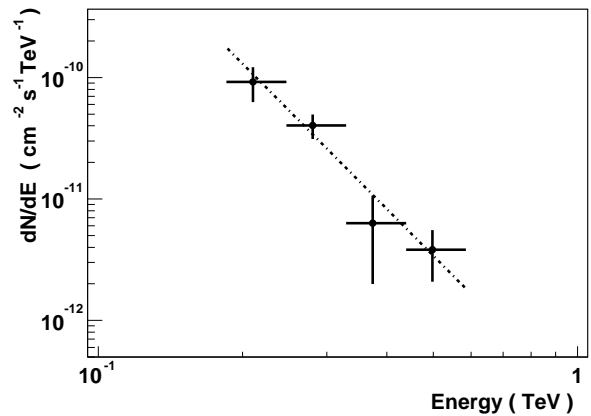} \\ [-0.3cm]
      \caption{The energy spectrum of PG\,1553+113.  The horizontal error bars indicate
	the energy bin size. The dashed line
	represents the best $\chi^2$ fit of a power law.}
         \label{spectrum_plot}
   \end{figure}

Given the low number of excess events found using the H.E.S.S. standard analysis, 
the determination of an energy spectrum is not possible.
Therefore, a set of lower-threshold cuts were generated to increase the selection
efficiency for low-energy $\gamma$-rays, allowing for a spectral determination.  
The cuts were generated
using the exact same procedure as for the standard analysis cuts, 
differing only in that they were optimized for a 1\% Crab 
flux ($>$100 GeV) source with photon index $\Gamma$=5.0.   Using these cuts 
(also shown in Table~\ref{thecuts}) an excess of 230 events ($5.3$$\sigma)$ 
from the direction of PG\,1553+113 is found.  
The significance of this excess is clearly 
subject to a small trials factor and is therefore reduced.  To be conservative,
only the significance of the standard analysis should be considered.
The energy spectrum determined from the data set, using the low-threshold analysis,
is shown in Figure~\ref{spectrum_plot}.  The best $\chi^2$ fit of a 
power law to these data yields a photon index $\Gamma$=4.0$\pm$$0.6_{stat}$, 
and a $\chi^2$ of 1.8 for 2 degrees of freedom.  
The systematic error on the photon index is small compared to the statistical error.

It should be noted that the integral flux (again assuming $\Gamma$=4.0) determined with
the ''spectrum'' cuts, 
I($>$200 GeV)=(5.8$\pm$$1.1_{stat}$$\pm$$1.2_{syst}$)$\times$10$^{-12}$ cm$^{-2}$\,s$^{-1}$,
is consistent with the flux obtained using the standard cuts.  
This indicates that the increased excess found with these cuts is indeed
expected and not the result of over-optimization or statistical fluctuations.
In addition, a cross check was performed on the same data as published
by H.E.S.S. on PKS\,2005$-$489 \cite{pks2005_paper}.  Here an analysis using the
spectral-determination cuts results in the same spectrum ($\Gamma$=4.0) 
as previously published, as well as the increase of the detected excess
by similar factor as for PG\,1553+113.  
Finally, an alternative analysis using an independent calibration
and two different event reconstruction methods (\cite{mathieu_model, 3D_model})
verifies the observed signal ($>$5$\sigma$ excess) and yields compatible spectra.
 
\section{Discussion}

  \begin{figure}
   \centering
      \includegraphics[width=8.7cm]{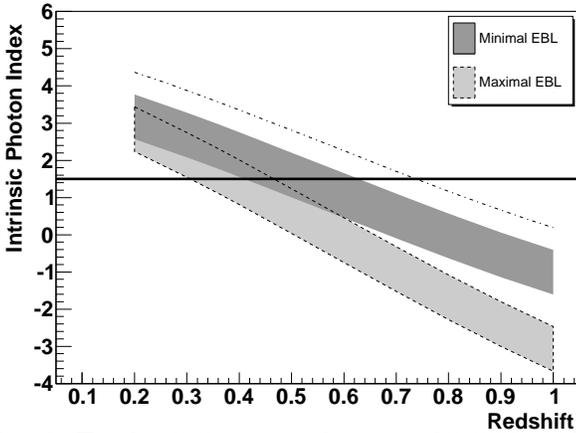} \\ [-0.3cm]
      \caption{The intrinsic photon index of PG\,1553+113, with 1$\sigma$ statistical error
contours, versus redshift for the cases of {\it minimal} and {\it maximal} EBL absorption.  
An intrinsic photon index below the horizontal line ($\Gamma_{\mathrm{int}}$=1.5) is not
considered realistic.  The uppermost (dashed-dotted) curve is the sum of the 
intrinsic photon index and 2$\sigma$ statistical error
for the case of minimal EBL absorption.}
         \label{index_vs_z_plot}
   \end{figure} 

PG\,1553+113 possibly represents the most distant BL Lac object detected at 
VHE energies. Therefore, the observed VHE spectrum is expected to be affected by EBL 
absorption. Unfortunately the unknown redshift of this object does not allow its 
VHE spectrum to be used in constraining the 
poorly-measured EBL flux density (see e.g. \cite{HESS_EBL_paper} 2005d).
However, assuming a redshift allows the
determination of the spectrum intrinsic to the BL Lac 
(i.e. with the effect of EBL absorption removed).
The intrinsic flux is related to the observed flux by 
F$_{\mathrm{int}}$\,=\,e$^{\tau(z,E)}$\,F$_{\mathrm{obs}}$,
where $\tau(z,E)$ is the optical depth due to pair production 
($\gamma_{_{\mathrm{VHE}}}$\,$\gamma_{_{\mathrm{EBL}}}$\,$\rightarrow$\,$e^{+}$\,$e^{-}$).  
Assuming that the main uncertainty arises from the overall normalization 
of the EBL density, this optical depth can be expressed as 
$\tau(z,E)$\,=\,$f$\,$\tau_{0}(z,E)$, where $\tau_{0}(z,E)$ is calculated
using a reference EBL-density model, and $f$ is a scaling factor
for this density. At energies relevant to the H.E.S.S. observations, 
$\tau(z,E)$ increases with both redshift and energy.  Therefore, a larger 
redshift, or higher EBL flux density, causes a stronger absorption 
of higher-energy $\gamma$-rays, resulting in a harder intrinsic spectrum 
(F$_{\mathrm{int}}$\,$\propto$\,$E^{-\Gamma_{\mathrm{int}}}$).
The EBL density model of \cite{joel_EBL} is chosen here since
it includes the effects of galaxy evolution which become
important at redshifts greater than $\sim$0.2. 
This model's fluxes are above recently published 
upper limits (\cite{HESS_EBL_paper} 2005d). However,
scaling its values by $f$=0.85 approximately traces these limits\footnote{A different model
and hence scaling factor is used in \cite{HESS_EBL_paper} (2005d).}
and is therefore considered a {\it maximal} EBL flux density.  
Given that the EBL flux density is poorly-known,
a {\it minimal} EBL scenario is also considered. Here, the original model is scaled by $f$=0.6, 
approximately tracing the lower limits on the EBL flux density 
provided by galaxy counts \cite{pozzetti}.  Using these parameterizations, the intrinsic spectrum of 
PG\,1553+113 is calculated for different redshift values.  Figure~\ref{index_vs_z_plot}
shows the intrinsic photon index ($\Gamma_{\mathrm{int}}$) versus redshift. 
Assuming the intrinsic photon index of the BL Lac
is not harder than $\Gamma_{\mathrm{int}}$=1.5 (see \cite{HESS_EBL_paper} 2005d for arguments 
supporting this assumption), an upper limit on the redshift of PG\,1553+113 of $z$$<$0.74 is 
determined using the {\it minimal} EBL.  
Above this redshift, where $\Gamma_{\mathrm{int}}$=0.3$\pm$0.6, 
the sum of $\Gamma_{\mathrm{int}}$ and twice its statistical error is harder than 1.5. 
Clearly, the redshift upper limit is smaller if a higher EBL flux density is assumed. 
This is the first time that a VHE observation is used to usefully constrain 
the redshift of an astrophysical object.  Future observations of PG\,1553+113
should allow for a more precise determination of the photon index, possibly
resulting in a stronger redshift constraint.

\begin{acknowledgements}
The support of the Namibian authorities and of the University of Namibia
in facilitating the construction and operation of H.E.S.S. is gratefully
acknowledged, as is the support by the German Ministry for Education and
Research (BMBF), the Max-Planck-Society, the French Ministry for Research,
the CNRS-IN2P3 and the Astroparticle Interdisciplinary Programme of the
CNRS, the U.K. Particle Physics and Astronomy Research Council (PPARC),
the IPNP of the Charles University, the South African Department of
Science and Technology and National Research Foundation, and by the
University of Namibia. We appreciate the excellent work of the technical
support staff in Berlin, Durham, Hamburg, Heidelberg, Palaiseau, Paris,
Saclay, and in Namibia in the construction and operation of the
equipment.
\end{acknowledgements}

\end{document}